\documentclass{appolb}
\usepackage{epsfig}
\usepackage[T1]{fontenc}
\begin{document}
\title{Liquid-Argon Time Projection Chambers in the U.S.%
\thanks{Presented at the 45th Winter School in Theoretical Physics ``Neutrino Interactions: from Theory to Monte Carlo Simulations'', L\k{a}dek-Zdr\'oj, Poland, February 2--11, 2009.}%
}
\author{Mitchell Soderberg
\address{Yale University}
}
\maketitle
\begin{abstract}
Liquid Argon Time Projection Chamber (LAr TPC) detectors are ideally suited for studying neutrino interactions and probing the parameters that characterize neutrino oscillations.    The ability to drift ionization particles over long distances in purified argon and to trigger on abundant scintillation light allows for excellent particle identification and triggering capability.  Recent U.S. based work in the development of LAr TPC technology for massive kiloton size detectors will be discussed in this talk, including details of the ArgoNeuT (Argon Neutrino Test) test-beam project, which is a 175 liter LAr TPC exposed to Fermilab's NuMI neutrino beamline.    
\end{abstract}
\PACS{PACS numbers come here}

\section{Introduction}
Liquid Argon Time Projection Chamber (LAr TPC) detectors are an intriguing option for future neutrino experiments that will measure the properties of neutrino oscillations.  The unique combination of position resolution, calorimetry, and scalability provided by LArTPCs make them a desirable technology choice for future massive detectors.  These detectors are not without their challenges, among which the requirement of very clean argon stands out.  With an interest in pursuing this technology for future massive detectors, several groups in the U.S. have been performing research and development of LArTPCs.  In this paper several of the components of this research will be briefly described. 

\section{LAr TPC Technique}
The LArTPC technique has been around for several decades, with pioneering work done as part of the ICARUS experiment \cite{Rubbia, ICARUS}.  A wire chamber is placed in highly-purified liquid argon, and an electric field is created within this detector.  Particle interactions with the argon inside the detector volume produce ionization electrons that drift along the electric field until they reach the wireplanes, upon which they will produce signals that are utilized for imaging purposes.  Applying proper bias voltages to the wireplanes allows several complementary views of the same interaction, providing the information necessary for three-dimensional event imaging\cite{Grids}.  This technique allows for very precise imaging, the resolution being dependent on several factors such as:  wire pitch, plane spacing, sampling rate, and electronics S/N levels.  The technology is attractive in that the number of electronics channels required for the detector does not scale directly with the volume of the detector.   This scaling feature, along with the relatively low cost of Argon, makes LArTPCs an intriguing option for future massive neutrino detectors.

One of the biggest challenges these detectors must address is producing Argon that is pure enough to allow the ionization to drift for the required distances.  Several U.S. institutions have worked on developing new filters that can cleanse the Argon to the required levels necessary for a LArTPC experiment, and can also be regenerated when they have become saturated \cite{Filter}.  These new filters are a necessary step along the path to massive detectors, and they also have allowed several teststands to be built in the U.S. with the goals of studying detector material effects on argon purity, and looking for cosmic-ray events in an LArTPC \cite{Yale}.

\section{ArgoNeuT}
\begin{figure}[htbp] 
   \centering
   \includegraphics[height=2.0in,width=3in]{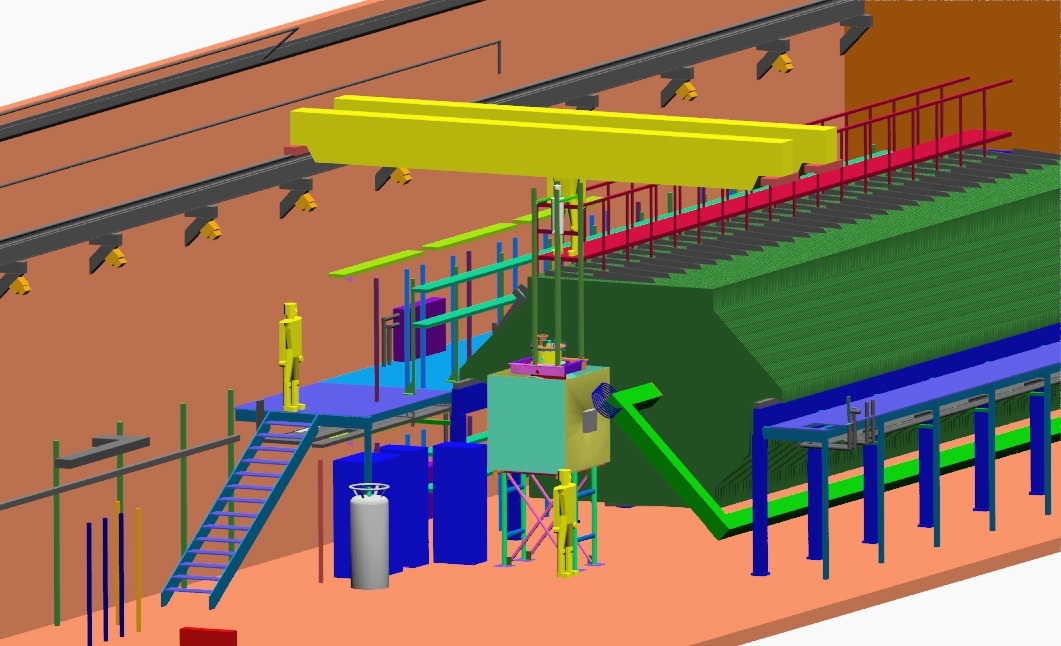} 
   \caption{Schematic diagram of ArgoNeuT location in NuMI tunnel.  The MINOS near detector is directly downstream of ArgoNeuT.}
   \label{fig:numi}
\end{figure}

\begin{figure}[htbp] 
   \centering
   \includegraphics[height=1.75in,width=2.25in]{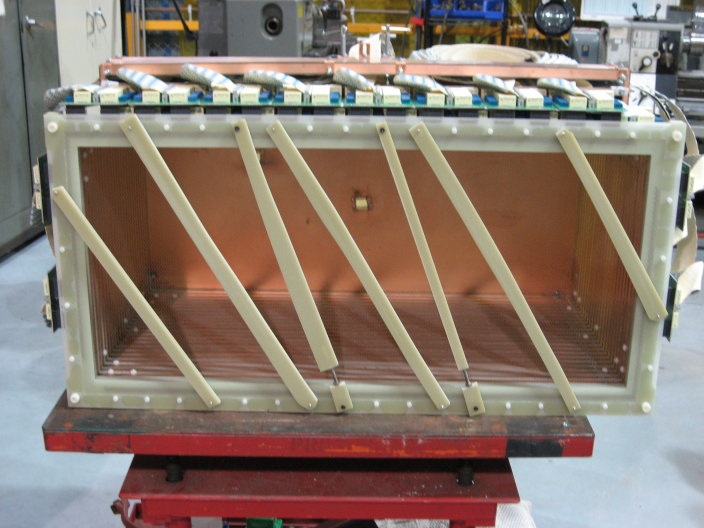} 
   \includegraphics[height=1.75in,width=2.25in]{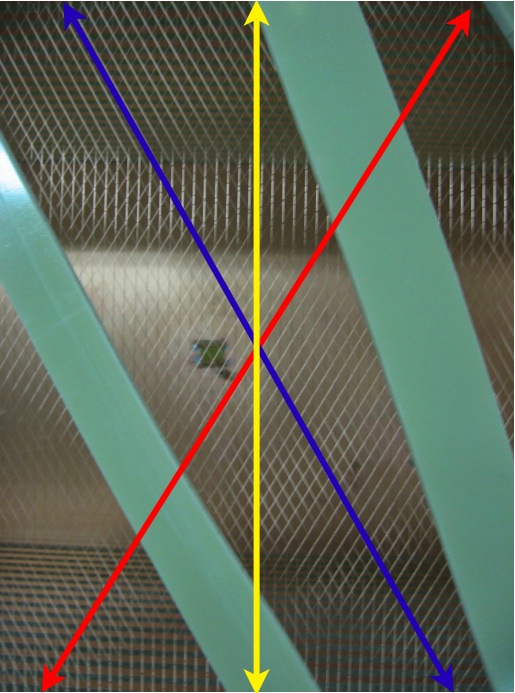} 
   \caption{Left:  The ArgoNeuT TPC.   Right:  Wire orientations of ArgoNeuT TPC (Collection=-60$^\circ$, Induction=+60$^\circ$)}
   \label{fig:tpc}
\end{figure}

Collecting a sample of neutrino interactions in Argon is one of the early focuses of the U.S. LArTPC program, since such an event sample will be very useful for developing the software tools necessary for data analysis.  The realization of this goal comes in the form of ArgoNeuT, a small test-beam project to put a LArTPC in a low-energy neutrino beam.  The detector is a 175-liter TPC housed in a 500-liter cryostat, positioned in the on-axis NuMI beamline at Fermilab.  As shown in Fig. \ref{fig:numi}, the detector is directly upstream of the MINOS near-detector, allowing muons exiting ArgoNeuT to range-out in MINOS and be recovered for analysis.  The TPC, shown in Fig. \ref{fig:tpc}, consists of 480 channels of electronic readout, distributed evenly between an induction plane and a collection plane.  There is also an inner ``shield" plane, which is not instrumented, that is used for electric field shaping.  ArgoNeuT uses a closed-loop cryogenic recirculation system to continually cleanse the liquid Argon to the desired purity levels.  ArgoNeuT had a $\approx$45 day commissioning run on the surface at Fermilab in the summer of 2008, during which many cosmic-ray muons were collected.  Several improvements were made to the detector following this commissioning run, and as of May 2009 the detector is operational in the NuMI tunnel and recording neutrino interactions.  ArgoNeuT will run through the end of 2009 in the NuMI beam, collecting several hundred neutrino interactions each day.  Work is already underway to develop software tools for ArgoNeuT, and also to produce simulated event samples which can be used in data analysis.

\subsection{ArgoNeuT Electronics}
ArgoNeuT uses new electronics, which is an important milestone for the development of LArTPCs in the U.S.  A FET preamplifier similar to the D0/ICARUS front-end is utilized.  A wide bandwith filter, from 10-200kHz, is applied to each channel.  The philosophy is to record as much information as possible, and employ digital signal processing (DSP) offline to extract hit/track parameters.  The electronics system has been built to minimize noise sources, including the preamplifiers being housed in a double-shielded Faraday cage which is cooled remotely via ducting.  In addition the preamplifier power supplies have extensive DC power filtering built in.

\section{MicroBooNE}
The next step in the U.S. LArTPC program is MicroBooNE, a proposed experiment that will help answer important physics questions and will also provide useful hardware development for future massive detectors.  MicroBooNE consists of a $\approx$90 ton TPC which will sit in a 175ton cryostat.  The TPC will contain 3 instrumented wireplanes with 3mm pitch, totaling approximately 10000 channels.  This experiment will sit on the surface at Fermilab, and will be simultaneously exposed to the on-axis Booster neutrino beamline, and the off-axis NuMI beamline.  MicroBooNE will utilize 30 cryogenic photomultiplier tubes that will provide a trigger which will be required in coincidence with a beam spill in order to reduce the data output of the experiment.
  
  \begin{figure}[htbp] 
   \centering
   \includegraphics[height=2.75in,width=3in]{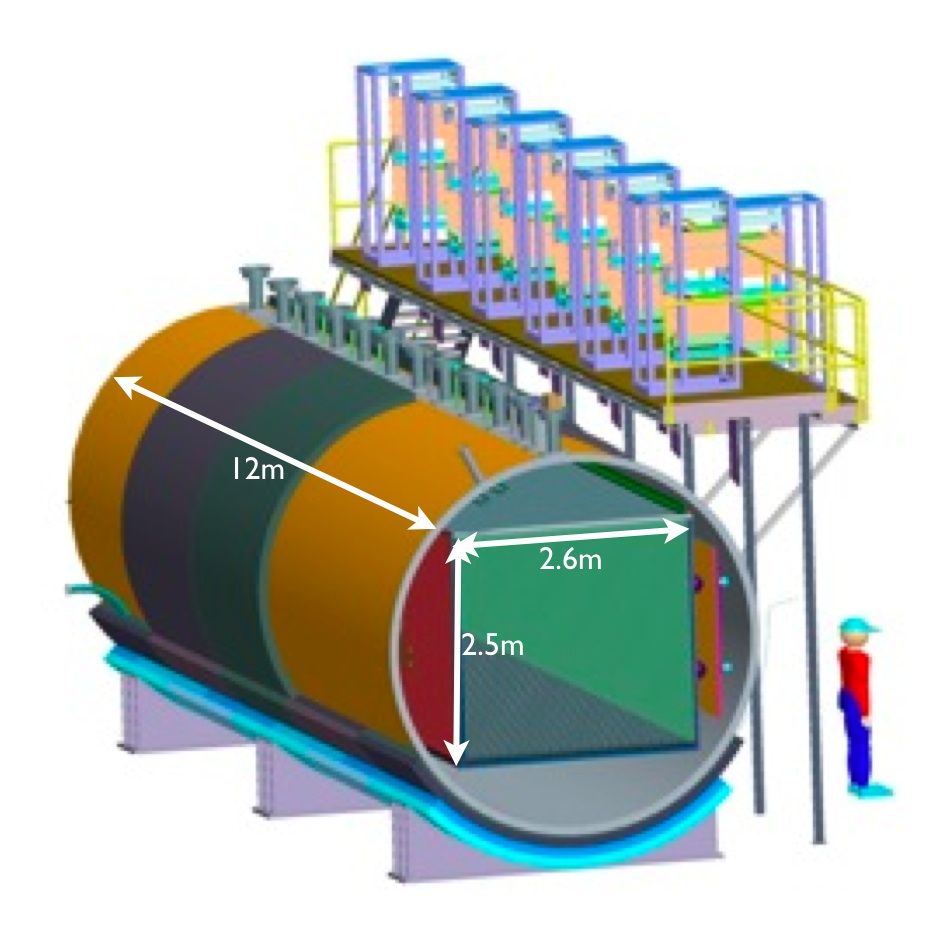}  
   \caption{Schematic diagram of MicroBooNE experiment.}
   \label{fig:microboone}
\end{figure}

MicroBooNE will have several features that will provide valuable data for future massive LArTPC detectors.  For example, MicroBooNE will place its preamplifier channels inside of the cryostat, in an 8$\%$ gas ullage at the top of the cryostat.  The cold temperature of the ullage gas should allow the preamplifiers to operate with a S/N a factor of 3 better than room temperature performance.  Placing the electronics inside the cryostat is likely to be a requirement for future massive detectors, which will have to do everything possible to combat noise incurred during the long transit a signal must make up and out of the very large detector.  By amplifying the signal close to the source, noise levels in the data can be kept to a minimum.   

Another interesting feature of the MicroBooNE experiment is a purity demonstration it will carry out, filling the cryostat with liquid after a gaseous purge instead of an initial vacuum pumpout.  The goal of this is to determine whether or not the required argon purity can be achieved starting from a non-evacuated environment.  This is crucial information needed in the design of massive detectors, which may be so large that they cannot be evacuated due to structural risks.  

\section{Massive Detectors}
Beyond MicroBooNE there is a desire to build a very massive ($\geq$5kton) LArTPC which would act as the far detector in a long-baseline neutrino experiment.  Such a scenario is envisioned for the Deep Underground Science and Engineering Laboratory (DUSEL) located in Homestake, South Dakota.  This facility offers the option of several different depths (100m, 1500m) at which a massive neutrino detector might be placed.  DUSEL would ideally be combined with an intense neutrino beam emanating from Fermilab, 1300km away.  The physics opportunities provided by the combination of the DUSEL depth and the Fermilab beam are quite extraordinary.  A massive LArTPC detector could perform neutrino oscillation physics using the Fermilab beam, while simultaneously performing proton-decay searches and watching for supernova neutrinos.

Placing a massive LArTPC at such a depth will be a major feat of engineering.  There are very serious safety considerations that must be addressed in order to mitigate the chance of any oxygen-deficiency situations arising from a leak.  Maintaining the desired level of argon purity for $\geq$5 ktons of liquid will also be a tremendous challenge.

\section{Conclusion}

There is a very active program of LArTPC development in the U.S., geared towards the goal of building a massive detector to study neutrino oscillation physics.  Several test projects already exist in the U.S. to study Argon purity, and to collect a sizable sample of neutrino interactions that can be utilized to develop software tools.  MicroBooNE is the next step in the U.S. for LArTPC development, and aside from performing interesting physics analyses, it will help answer several hardware questions that are important for future massive detectors.

\end{document}